\documentclass[letterpaper,twocolumn,american,showpacs,pra,aps,superscriptaddress]{revtex4}
\usepackage[latin1]{inputenc}
\usepackage{bm}
\usepackage{amsmath}
\usepackage{graphicx}
\usepackage{amssymb}
\usepackage{hyperref}
\makeatletter
\usepackage{pifont}
\makeatother

\begin{document}

\title{Separability criteria and bounds for entanglement measures}

\author{Heinz-Peter Breuer}

\email{breuer@physik.uni-freiburg.de}

\affiliation{Physikalisches Institut, Universit\"at Freiburg,
             Hermann-Herder-Str.~3, D-79104 Freiburg, Germany}

\date{\today}

\begin{abstract}
Employing a recently proposed separability criterion we develop
analytical lower bounds for the concurrence and for the
entanglement of formation of bipartite quantum systems. The
separability criterion is based on a nondecomposable positive map
which operates on state spaces with even dimension $N\geq 4$, and
leads to a class of nondecomposable optimal entanglement
witnesses. It is shown that the bounds derived here complement and
improve the existing bounds obtained from the criterion of
positive partial transposition and from the realignment criterion.
\end{abstract}


\maketitle

\section{Introduction}

A central problem in quantum information theory
\cite{ALBER,ECKERT} is the formulation of appropriate measures
that quantify the degree of entanglement in composite systems.
Particularly important entanglement measures are the concurrence
\cite{WOOTTERS1,WOOTTERS2,RUNGTA} and the entanglement of
formation \cite{BENNETT,VIDAL}. These quantities have been widely
used in many applications. Examples include studies on the role of
entanglement in quantum phase transitions
\cite{OSTERLOH,OSBORNE,LIDAR}, on the emergence of long-distance
entanglement in spin systems \cite{VENUTI}, and on additivity
properties of the Holevo capacity of quantum channels \cite{SHOR}.

The explicit determination of most of the proposed entanglement
measures for a generic state $\rho$ is an extremely demanding task
that requires the solution of a high-dimensional optimization
problem. The development of analytical lower bounds for the
various entanglement measures is therefore of great interest.
Recently, Chen, Albeverio, and Fei \cite{ALBEVERIO1,ALBEVERIO2}
have derived such bounds for the concurrence $C(\rho)$ and for the
entanglement of formation $E(\rho)$. They achieved this by
relating $C(\rho)$ and $E(\rho)$ to two important and strong
separability criteria, namely to the Peres criterion of positive
partial transposition (PPT) \cite{PERES,HORODECKI96} and to the
realignment criterion \cite{CHEN,RUDOLPH}. According to these
criteria a given state $\rho$ is entangled (inseparable) if the
trace norms $||T_2\rho||$ or $||{\mathcal{R}}\rho||$ are strictly
larger than 1, where $T_2$ denotes the partial transposition and
${\mathcal{R}}$ the realignment transformation. In
Refs.~\cite{ALBEVERIO1,ALBEVERIO2} tight lower bounds for
$C(\rho)$ and $E(\rho)$ have been formulated in terms of these
trace norms.

Here, we extend the connection between separability criteria and
entanglement measures to a new criterion which has been developed
recently \cite{TRC-PAPER}. This criterion is based on a universal
nondecomposable positive map which leads to a class of optimal
entanglement witnesses. Employing these witnesses we derive
analytical lower bounds for the concurrence and for the
entanglement of formation that can be expressed in terms of a
simple linear functional of the given state $\rho$.

The entanglement witnesses constructed here have the special
feature of being nondecomposable optimal. This notion has been
introduced in Refs.~\cite{LEWENSTEIN00,LEWENSTEIN01} to
characterize optimality properties of entanglement witnesses. It
means that the witnesses are able to identify entangled PPT states
and that no other witnesses exist which can detect more such
states. It follows that the bounds developed here can be sharper
than those obtained from the PPT criterion and that they are
particularly efficient near the border that separates the PPT
entangled states from the separable states. In addition, we will
demonstrate that they can also be stronger than the bounds given
by the realignment criterion. Hence, the new bounds complement and
considerably improve the existing bounds.

The paper is organized as follows. In Sec.~\ref{MAPS} we introduce
a new separability criterion which is based on a nondecomposable
positive map that operates on the states of Hilbert spaces with
even dimension $N \geq 4$. We formulate and prove the most
important properties of this map, and derive the associated class
of optimal entanglement witnesses. Analytical lower bounds for the
concurrence are developed in Sec.~\ref{SEC-CONCURRENCE}. In
Sec.~\ref{SEC-EXAMPLE} we discuss an example of a certain family
of states in arbitrary dimensions. It will be demonstrated
explicitly with the help of this example that the new bounds can
be much sharper than the bounds of the PPT and of the realignment
criterion. The new class of entanglement witnesses is used in
Sec.~\ref{SEC-EOF} to develop corresponding lower bounds for the
entanglement of formation. Finally, some conclusions are drawn in
Sec.~\ref{CONCLU}.

\section{Separability criteria}\label{MAPS}

We consider a quantum system with finite-dimensional Hilbert space
${\mathbb C}^N$. Without loss of generality one can regard
${\mathbb C}^N$ as the state space of a particle with a certain
spin $j$, where $N=2j+1$. As usual, the corresponding basis states
are denoted by $|j,m\rangle$, where the quantum number $m$ takes
on the values $m=-j,-j+1,\ldots,+j$.

\subsection{Time reversal transformation}\label{TIME-REVERSAL}

We will develop a necessary condition for the separability of
mixed quantum states which employs the symmetry transformation of
the time reversal \cite{GALINDO}. In quantum mechanics the time
reversal is to be described by an antiunitary operator $\theta$.
As for any antiunitary operator, we can write $\theta=V\theta_0$,
where $\theta_0$ denotes the complex conjugation in the chosen
basis $|j,m\rangle$, and $V$ is a unitary operator. In the spin
representation introduced above the matrix elements of $V$ are
given by $\langle j,m'|V|j,m\rangle=(-1)^{j-m}\delta_{m',-m}$. For
even $N$, i.~e., for half-integer spins $j$, this matrix is not
only unitary but also skew-symmetric, which means that $V^T=-V$,
where $T$ denotes the transposition. It follows that $\theta^2=-I$
which leads to
\begin{equation} \label{ORTHO}
 \langle\varphi|\theta\varphi\rangle = 0.
\end{equation}
This relation expresses a well-known property of the time reversal
transformation $\theta$ which will play a crucial role in the
following: For any state vector $|\varphi\rangle$ the
time-reversed state $|\theta\varphi\rangle$ is orthogonal to
$|\varphi\rangle$. This is a distinguished feature of
even-dimensional state spaces, because unitary and skew-symmetric
matrices do not exist in state spaces with odd dimension.

The action of the time reversal transformation on an operator $B$
on ${\mathbb C}^N$ can be expressed by
\begin{equation} \label{DEF-THETA}
 \vartheta B = \theta B^{\dagger} \theta^{-1}
 = V B^T V^{\dagger}.
\end{equation}
This defines a linear map $\vartheta$ which transforms any
operator $B$ to its time reversed operator $\vartheta B$. For
example, if we take the spin operator $\hat{\bm{j}}$ of the
spin-$j$ particle this gives the spin flip transformation
$\vartheta \hat{\bm{j}} = -\hat{\bm{j}}$.

\subsection{Nondecomposable positive maps and optimal entanglement witnesses}

A positive map is a linear transformation $\Lambda$ which takes
any positive operator $B$ on some state space ${\mathcal{H}}_2$ to
a positive operator $\Lambda B$, i.~e., $B \geq 0$ implies that
$\Lambda B \geq 0$. A positive map $\Lambda$ is said to be
completely positive if it has the additional property that the map
$I\otimes \Lambda$ operating on any composite system with state
space ${\mathcal{H}}_1\otimes{\mathcal{H}}_2$ is again positive,
where $I$ denotes the unit map. The physical significance of
positive maps in entanglement theory is provided by a fundamental
theorem established in Ref.~\cite{HORODECKI96}. According to this
theorem a necessary and sufficient condition for a state $\rho$ to
be separable is that the operator $(I\otimes \Lambda)\rho$ is
positive for any positive map $\Lambda$. Hence, maps which are
positive but not completely positive can be used as indicators for
the entanglement of certain sets of states.

An important example for a positive but not completely positive
map is given by the transposition map $T$. The inequality $T_2\rho
\equiv (I\otimes T)\rho\geq 0$ represents a strong necessary
condition for separability known as the Peres criterion of
positive partial transposition (PPT criterion). The second
relation of Eq.~(\ref{DEF-THETA}) shows that the time reversal
transformation $\vartheta$ is unitarily equivalent to the
transposition map $T$. Hence, the PPT criterion is equivalent to
the condition that the partial time reversal $\vartheta_2$ is
positive:
\[
 \vartheta_2 \rho \equiv (I \otimes \vartheta) \rho \geq 0.
\]

We define a linear map $\Phi$ which acts on operators $B$ on
${\mathbb C}^N$ as follows \cite{TRC-PAPER}
\begin{equation} \label{PHI}
 \Phi B = ({\mathrm{tr}}B) I - B - \vartheta B,
\end{equation}
where ${\mathrm{tr}}B$ denotes the trace of $B$ and $I$ is the
unit operator. This map has first been introduced in
Ref.~\cite{NtensorN} for the special case $N=4$, in order to study
the entanglement structure of SU(2)-invariant spin systems
\cite{SU2,SCHLIEMANN}.

For any even $N \geq 4$ the map $\Phi$ defined by Eq.~(\ref{PHI})
has the following features:
\begin{description}
 \item[(A)] $\Phi$ is a positive but not completely positive map.
 \item[(B)] The map $\Phi$ is nondecomposable.
 \item[(C)] The entanglement witnesses corresponding to $\Phi$ are
           nondecomposable optimal.
\end{description}
In the following we briefly explain and prove these statements.

\textbf{(A)} We first demonstrate that $\Phi$ is a positive map.
To this end, we have to show that the operator
$\Phi(|\varphi\rangle\langle\varphi|)$ is positive for any
normalized state vector $|\varphi\rangle$. Using definition
(\ref{PHI}) we find:
\[
 \Phi(|\varphi\rangle\langle\varphi|)
 = I - |\varphi\rangle\langle\varphi|
 - |\theta\varphi\rangle\langle\theta\varphi| \equiv I - \Pi.
\]
Because of Eq.~(\ref{ORTHO}) the operator $\Pi$ introduced here
represents an orthogonal projection operator which projects onto
the subspace spanned by $|\varphi\rangle$ and
$|\theta\varphi\rangle$. It follows that also
$\Phi(|\varphi\rangle\langle\varphi|)$ is a projection operator
and, hence, that it is positive for any normalized state vector
$|\varphi\rangle$. This proves that $\Phi$ is a positive map.

We remark that for $N=2$ the projection $\Pi$ is identical to the
unit operator such that $\Phi$ is equal to the zero map in this
case. For this reason we restrict ourselves to the cases of even
$N\geq 4$. It should be emphasized that $\Phi$ would not be
positive if we had used the transposition $T$ instead of the time
reversal $\vartheta$ in the definition (\ref{PHI}).

The positivity of $\Phi$ implies that the inequality
\begin{equation} \label{TRC}
 \Phi_2 \rho \equiv (I \otimes \Phi) \rho \geq 0
\end{equation}
provides a necessary condition for separability: any state $\rho$
which violates this condition must be entangled. To show that
$\Phi$ is not completely positive, i.~e., that the condition
(\ref{TRC}) is nontrivial, we consider the tensor product space
${\mathcal{H}}_1\otimes{\mathcal{H}}_2={\mathbb
C}^N\otimes{\mathbb C}^N$ of two spin-$j$ particles. The total
spin of the composite system will be denoted by $J$. According to
the triangular inequality $J$ takes on the values
$J=0,1,\ldots,2j=N-1$. The projection operator which projects onto
the manifold of states corresponding to a definite value of $J$
will be denoted by $P_J$. In particular, $P_0$ represents the
one-dimensional projection onto the maximally entangled singlet
state $J=0$.

We define a Hermitian operator $W$ by applying $I\otimes \Phi$ to
the singlet state:
\begin{equation} \label{DEF-W}
 W \equiv N(I \otimes \Phi) P_0,
\end{equation}
where the factor $N$ is introduced for convenience. More explicit
expressions for $W$ can be obtained as follows. First, we observe
that ${\mathrm{tr}}_2 P_0 = I/N$ since $P_0$ is a maximally
entangled state (${\mathrm{tr}}_2$ denotes the partial trace taken
over subsystem 2). Second, we note that the partial time reversal
of the singlet state is given by the formula \cite{NtensorN}
\begin{equation} \label{THETA-P0}
 \vartheta_2P_0 = \frac{1}{N} F = -\frac{1}{N}\sum_{J=0}^{2j} (-1)^JP_J,
\end{equation}
where $F$ denotes the swap operator defined by
\begin{equation} \label{SWAP}
 F|\varphi_1\rangle \otimes |\varphi_2\rangle = |\varphi_2\rangle
\otimes |\varphi_1\rangle.
\end{equation}
Using then definition (\ref{PHI}) we get
\begin{equation} \label{DEF-W-2}
 W = I - NP_0 - F.
\end{equation}
Another useful representation is obtained by use of the fact that
the sum of the $P_J$ is equal to the unit operator. Expressing $F$
as shown in Eq.~(\ref{THETA-P0}) we then find:
\begin{equation} \label{DEF-W-3}
 W = -(N-2)P_0 + 2P_2 + 2P_4 + \ldots + 2P_{2j-1}.
\end{equation}
We infer from Eq.~(\ref{DEF-W-3}) that $W$ has the negative
eigenvalue $-(N-2)$ corresponding to the singlet state $J=0$.
Therefore, the operator $W$ is not positive and, hence, the map
$\Phi$ is not completely positive.

\textbf{(B)} Since $\Phi$ is positive but not completely positive
the operator $W$ defined in Eq.~(\ref{DEF-W}) is an entanglement
witness \cite{HORODECKI96,TERHAL}. We recall that an entanglement
witness is an observable which satisfies
${\mathrm{tr}}\{W\sigma\}\geq 0$ for all separable states
$\sigma$, and ${\mathrm{tr}}\{W\rho\}< 0$ for at least one
inseparable state $\rho$, in which case we say that $W$ detects
$\rho$.

An entanglement witness $W$ is called nondecomposable if it can
detect entangled PPT states \cite{LEWENSTEIN00}, i.~e., if there
exist PPT states $\rho$ that satisfy ${\mathrm{tr}}\{ W \rho \} <
0$. We will demonstrate in Sec.~\ref{SEC-EXAMPLE} by means of an
explicit example that there are always such states for the witness
defined by Eq.~(\ref{DEF-W}). It follows that our witness $W$ is
nondecomposable. This implies that also the map $\Phi$ is
nondecomposable \cite{WORONOWICZ}, and that the criterion
(\ref{TRC}) is able to detect entangled PPT states.

\textbf{(C)} The observable $W$ introduced above has a further
remarkable optimality property. To explain this property we
introduce the following notation \cite{LEWENSTEIN00}. We denote by
$D_{W}$ the set of all entangled PPT states of the total state
space which are detected by some given nondecomposable witness
$W$. A witness $W_2$ is said to be finer than a witness $W_1$ if
$D_{W_1}$ is a subset of $D_{W_2}$, i.~e., if all entangled PPT
states which are detected by $W_1$ are also detected by $W_2$. A
given witness is said to be nondecomposable optimal if there is no
other witness which is finer, i.~e., if there is no other witness
which is able to detect more entangled PPT states.

It can be shown that the witness $W$ defined by (\ref{DEF-W}) is
always optimal in this sense. The proof can be carried out by
showing that the set of product vectors
$|\varphi_1,\varphi_2\rangle$ satisfying
$\langle\varphi_1,\varphi_2 |W|\varphi_1,\varphi_2\rangle = 0$
spans the total Hilbert space. The details of the proof are given
in Ref.~\cite{TRC-PAPER}.

\section{Bounds for the concurrence}\label{SEC-CONCURRENCE}

The generalized concurrence of a pure state
$\rho=|\psi\rangle\langle\psi|$ is defined by \cite{RUNGTA}
\[
 C(|\psi\rangle) = \sqrt{2(1-{\mathrm{tr}}_1 \rho_1^2)},
\]
where $\rho_1 = {\mathrm{tr}}_2 \rho$ represents the reduced
density matrix of subsystem 1, given by the partial trace taken
over subsystem 2. We consider the Schmidt decomposition
\begin{equation} \label{SCHMIDT}
 |\psi\rangle = \sum_i \alpha_i |\varphi_i\rangle \otimes
 |\chi_i\rangle,
\end{equation}
where $\{|\varphi_i\rangle\}$ and $\{|\chi_i\rangle\}$ are
orthonormal bases in ${\mathcal{H}}_1$ and ${\mathcal{H}}_2$,
respectively, and the $\alpha_i$ are the Schmidt coefficients
satisfying $\alpha_i \geq 0$ and the normalization condition
\begin{equation} \label{NORMALIZATION}
 \sum_i \alpha_i^2 = 1.
\end{equation}
The concurrence can then be expressed in terms of the Schmidt
coefficients:
\begin{equation} \label{CONCU-PURE}
 C(|\psi\rangle) = \sqrt{2\sum_{i\neq j}\alpha_i^2\alpha_j^2}.
\end{equation}
For a mixed state $\rho$ the concurrence is defined to be
\begin{equation} \label{CONCURRENCE}
 C(\rho) = {\mathrm{min}} \left\{ \sum_r p_r C(|\psi_r\rangle)
 \; \Big| \; \rho = \sum_r p_r |\psi_r\rangle\langle\psi_r| \right\},
\end{equation}
where the minimum is taken over all possible convex decompositions
of $\rho$ into an ensemble $\{|\psi_r\rangle\}$ of pure states
with probability distribution $\{p_r\}$.

Let $\rho = \sum_r p_r |\psi_r\rangle\langle\psi_r|$ be an optimal
decomposition of $\rho$ for which the minimum of
Eq.~(\ref{CONCURRENCE}) is attained. Denoting the Schmidt
coefficients of $|\psi_r\rangle$ by $\alpha_i^r$ we then have:
\begin{eqnarray} \label{CONCU-1}
 C(\rho) &=& \sum_r p_r C(|\psi_r\rangle) \nonumber \\
 &=& \sum_r p_r \sqrt{2\sum_{i\neq j} (\alpha_i^r)^2
 (\alpha_j^r)^2} \nonumber \\
 &\geq& \sum_r p_r
 \sqrt{\frac{2}{N(N-1)}}
 \sum_{i \neq j} \alpha_i^r \alpha_j^r.
\end{eqnarray}
In the second line we have used Eq.~(\ref{CONCU-PURE}), and the
third line is obtained with the help of the inequality
\[
 \sum_{i \neq j} \alpha_i^2 \alpha_j^2 \geq
 \frac{1}{N(N-1)}
 \left( \sum_{i \neq j} \alpha_i \alpha_j \right)^2,
\]
which holds for any set of $N$ numbers $\alpha_i$
\cite{ALBEVERIO1}.

Consider now any real-valued and convex functional $f(\rho)$ on
the total state space with the following property. For all state
vectors $|\psi\rangle$ with Schmidt decomposition (\ref{SCHMIDT})
we have:
\begin{equation} \label{PROP-F}
 f(|\psi\rangle\langle\psi|) \leq \sum_{i \neq j} \alpha_i \alpha_j.
\end{equation}
Given such a functional we can continue inequality (\ref{CONCU-1})
as follows:
\begin{eqnarray*}
 C(\rho) &\geq& \sqrt{2/N(N-1)} \sum_r p_r
 \sum_{i \neq j} \alpha_i^r \alpha_j^r \\
 &\geq& \sqrt{2/N(N-1)} \sum_r p_r
 f(|\psi_r\rangle\langle\psi_r|) \\
 &\geq& \sqrt{2/N(N-1)} f\left( \sum_r p_r |\psi_r\rangle\langle\psi_r|\right) \\
 &=& \sqrt{2/N(N-1)} f(\rho).
\end{eqnarray*}
In the second line we have used inequality (\ref{PROP-F}), and in
the third line the convexity of $f(\rho)$.

We conclude that any convex functional $f(\rho)$ with the property
(\ref{PROP-F}) leads to a lower bound for the concurrence:
\[
 C(\rho) \geq \sqrt{\frac{2}{N(N-1)}} f(\rho).
\]
In Ref.~\cite{ALBEVERIO1} two example for such a functional
$f(\rho)$ have been constructed which are based on the PPT
criterion and on the realignment criterion:
\begin{eqnarray*}
 f_{\mathrm{ppt}}(\rho) &=& ||T_2\rho|| - 1, \\
 f_{\mathrm{realign}}(\rho) &=& ||{\mathcal{R}}\rho|| - 1,
\end{eqnarray*}
where $T_2$ denotes the partial transposition and ${\mathcal{R}}$
the realignment transformation. These functionals are convex
because of the convexity of the trace norm which is defined by
$||A||={\mathrm{tr}}\sqrt{A^{\dagger}A}$. Moreover, for both
functionals the equality sign of Eq.~(\ref{PROP-F}) holds:
\begin{equation} \label{PPT-REALIGN}
 f_{\mathrm{ppt}}(|\psi\rangle\langle\psi|)
 = f_{\mathrm{realign}}(|\psi\rangle\langle\psi|)
 = \sum_{i \neq j} \alpha_i \alpha_j.
\end{equation}

Consider the functional
\[
 f_W(\rho) = - {\mathrm{tr}}(W\rho),
\]
where $W$ is the entanglement witness introduced in
Eq.~(\ref{DEF-W}). This functional is linear and of course convex.
We claim that $f_W(\rho)$ also satisfies the bound (\ref{PROP-F}),
i.~e., for any state vector $|\psi\rangle$ with Schmidt
decomposition (\ref{SCHMIDT}) we have
\begin{equation} \label{INEQ-FW}
 f_W(|\psi\rangle\langle\psi|) \equiv -\langle\psi|W|\psi\rangle
 \leq \sum_{i \neq j} \alpha_i \alpha_j.
\end{equation}
To show this we first determine the expectation value of $W$. From
Eq.~(\ref{THETA-P0}) we have $NP_0=\vartheta_2 F$ and, hence, the
expression (\ref{DEF-W-2}) can be written as $W = I - F -
\vartheta_2 F$. This gives
\[
 \langle\psi|W|\psi\rangle
 = 1 - \langle\psi|F|\psi\rangle - \langle\psi|\vartheta_2
 F|\psi\rangle.
\]
With the help of the definitions of the swap operator $F$
[Eq.~(\ref{SWAP})] and of the time reversal transformation
[Eq.~(\ref{DEF-THETA})] it is easy to verify the formulae
\begin{eqnarray*}
 \langle\psi|F|\psi\rangle
 &=& \sum_{ij} \alpha_i \alpha_j
 \langle\varphi_i|\chi_j\rangle \langle\chi_i|\varphi_j\rangle, \\
 \langle\psi|\vartheta_2 F|\psi\rangle
 &=& \sum_{ij} \alpha_i \alpha_j
 \langle\varphi_i|\theta\chi_i\rangle
 \langle\theta\chi_j|\varphi_j\rangle.
\end{eqnarray*}
This leads to
\[
 \langle\psi|W|\psi\rangle
 = 1 - \sum_{ij} \alpha_i \alpha_j A_{ij},
\]
where
\begin{equation} \label{DEF-AIJ}
 A_{ij} \equiv \langle\varphi_i|\chi_j\rangle \langle\chi_i|\varphi_j\rangle +
 \langle\varphi_i|\theta\chi_i\rangle \langle\theta\chi_j|\varphi_j\rangle.
\end{equation}
Hence, we have
\[
 f_W(|\psi\rangle\langle\psi|) =
 \sum_{ij} \alpha_i \alpha_j A_{ij} - 1
 \leq \sum_{ij} \alpha_i \alpha_j |A_{ij}|  - 1.
\]
It is shown in Appendix \ref{APP-A} that
\begin{equation} \label{INEQ-AIJ}
 |A_{ij}| \leq 1.
\end{equation}
This leads immediately to the desired inequality:
\[
 f_W(|\psi\rangle\langle\psi|) \leq \sum_{ij} \alpha_i \alpha_j  - 1
 = \sum_{i \neq j} \alpha_i \alpha_j,
\]
where we have used the normalization condition
(\ref{NORMALIZATION}).

Summarizing we have obtained the following lower bound for the
concurrence:
\begin{equation} \label{NEWBOUND-CONCU}
 C(\rho) \geq -\sqrt{\frac{2}{N(N-1)}} {\mathrm{tr}}(W\rho).
\end{equation}
Of course, this bound is only nontrivial if $\rho$ is detected by
the entanglement witness $W$, i.~e., if ${\mathrm{tr}}(W\rho)<0$.
It will be demonstrated in Sec.~\ref{SEC-EXAMPLE} that this bound
can be much stronger than the bounds given by $f_{\mathrm{ppt}}$
and $f_{\mathrm{realign}}$, which is due to the fact $W$
identifies many entangled states that are neither detected by the
PPT criterion nor by the realignment criterion.

\section{Example}\label{SEC-EXAMPLE}
We illustrate the application of the inequality
(\ref{NEWBOUND-CONCU}) with the help of a certain family of
states. This family contains a separable state, entangled PPT
states, as well as entangled states whose partial transposition is
not positive. The example will also lead to a proof of the claim
that the map $\Phi$ and the witness $W$ are nondecomposable.

Consider the following one-parameter family of states:
\begin{equation} \label{FAMILY}
 \rho(\lambda) = \lambda P_0 + (1-\lambda) \rho_0,
 \qquad 0 \leq \lambda \leq 1.
\end{equation}
These normalized states are mixtures of the singlet state $P_0$
and of the state
\[
 \rho_0 = \frac{2}{N(N+1)} P_S = \frac{2}{N(N+1)} \sum_{J \; \mathrm{odd}}
 P_J,
\]
where $P_S$ denotes the projection onto the symmetric subspace
under the swap operation $F$. We note that $\rho_0$ is a separable
state which belongs to the class of the Werner states
\cite{WERNER}. Since $P_S$ can be written as a sum over the
projections $P_J$ with odd $J$, we immediately get with the help
of Eq.~(\ref{DEF-W}):
\begin{equation} \label{W-RHO}
 {\mathrm{tr}}(W\rho(\lambda)) = -\lambda(N-2).
\end{equation}
Hence, we find that ${\mathrm{tr}}(W \rho(\lambda)) < 0$ for
$\lambda > 0$. It follows that all states of the family
(\ref{FAMILY}) corresponding to $\lambda > 0$ are entangled, and
that $\rho_0$ is the only separable state of this family.

Employing Eqs.~(\ref{W-RHO}) and (\ref{NEWBOUND-CONCU}) we get the
following lower bound for the concurrence:
\begin{equation} \label{NEW-BOUND}
 C(\rho(\lambda)) \geq \sqrt{\frac{2(N-1)}{N}} \frac{N-2}{N-1}
 \lambda.
\end{equation}
To compare this bound with those obtained from the PPT and the
realignment criterion we have to determine the trace norms
$||T_2\rho(\lambda)||$ and $||{\mathcal{R}}\rho(\lambda)||$. The
details of the calculation are presented in Appendix \ref{APP-B}.
One finds that the PPT criterion gives the bounds:
\begin{eqnarray} \label{PPT-BOUND}
 \lefteqn{ C(\rho(\lambda)) \geq } \\
 && \left\{
 \begin{array}{ll}
 0, & \lambda \leq \frac{1}{N+2} \\
 \sqrt{\frac{2(N-1)}{N}} \frac{N-2}{N(N-1)} \left[ (N+2)\lambda - 1 \right],
 & \frac{1}{N+2} \leq \lambda \leq \frac{1}{2} \\
 \sqrt{\frac{2(N-1)}{N}} \frac{N\lambda-1}{N-1}, & \frac{1}{2} \leq \lambda
 \end{array} \right. \nonumber
\end{eqnarray}
while the realignment criterion yields:
\begin{equation} \label{REALIGN-BOUND}
 C(\rho(\lambda)) \geq \left\{
 \begin{array}{ll}
 \sqrt{\frac{2(N-1)}{N}} \frac{-2\lambda}{N-1},
 & \lambda \leq \frac{1}{N+2} \\
 \sqrt{\frac{2(N-1)}{N}} \frac{N\lambda-1}{N-1}, &
 \frac{1}{N+2} \leq \lambda
 \end{array} \right.
\end{equation}

The relations (\ref{NEW-BOUND})-(\ref{REALIGN-BOUND}) lead to a
number of important conclusions. First of all, we observe from
Eq.~(\ref{PPT-BOUND}) that the states within the range $\lambda
\leq 1/(N+2)$ have positive partial transposition (in this range
$||T_2\rho(\lambda)||$ is equal to $1$, see Appendix \ref{APP-B}).
But from Eq.~(\ref{W-RHO}) we know that all states with $\lambda >
0$ must be entangled. It follows that all states in the range $0 <
\lambda \leq 1/(N+2)$ are entangled PPT states which are detected
by the witness $W$. This proves, as claimed in Sec.~\ref{MAPS},
that the witness $W$ and, hence, also the map $\Phi$ are
nondecomposable.

According to Eq.~(\ref{PPT-BOUND}) the PPT criterion only detects
the entanglement of the states with $\lambda > 1/(N+2)$. It is
thus weaker than the criterion based on the witness $W$. As can be
seen from Eq.~(\ref{REALIGN-BOUND}) the realignment criterion is
even weaker because it only recognizes the entanglement of the
states with $\lambda > 1/N$ (the trace norm of
${\mathcal{R}}\rho(\lambda)$ is larger than $1$ if and only if
$\lambda > 1/N$, see Appendix \ref{APP-B}).

\begin{figure}[htb]
\includegraphics[width=\linewidth]{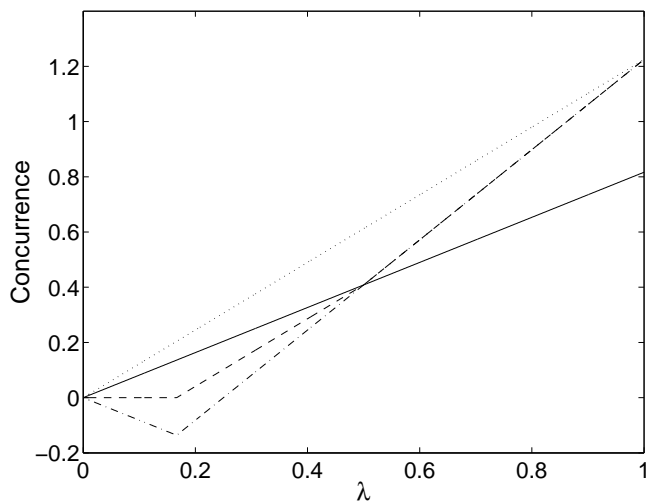}
\caption{Concurrence of the states (\ref{FAMILY}) for $N=4$. Solid
line: The new lower bound given by Eq.~(\ref{NEW-BOUND}). Dashed
line: Lower bound given by the PPT criterion
[Eq.~(\ref{PPT-BOUND})]. Dashed-dotted line: Lower bound given by
the realignment criterion [Eq.~(\ref{REALIGN-BOUND})]. Dotted
line: Upper bound given by $C(\rho(\lambda))\leq
\sqrt{2(N-1)/N}\lambda$.\label{figure1}}
\end{figure}

A plot of the various lower bounds for the example $N=4$ is shown
in Fig.~\ref{figure1}. We see that the new bound (\ref{NEW-BOUND})
is the best one within the range $\lambda < 1/2$. The bounds given
by the PPT and the realignment criterion coincide in the range
$\lambda > 1/2$. In this range they are better than the new bound.
Note that these features hold true for all $N$. We remark that for
large $N$ the concurrence approaches the limit $C(\rho(\lambda)) =
\sqrt{2}\lambda$.

\section{Entanglement of formation}\label{SEC-EOF}

For a pure state $|\psi\rangle$ with Schmidt decomposition
(\ref{SCHMIDT}) one defines the entanglement of formation by
\[
 E(|\psi\rangle) = H(\bm{\alpha}) \equiv -\sum_i
 \alpha_i^2 \log \alpha_i^2,
\]
where $\bm{\alpha}$ denotes the vector of the Schmidt
coefficients, and $\log$ is the base $2$ logarithm. The quantity
$H(\bm{\alpha})$ is the Shannon entropy of the distribution
$\alpha_i^2$, which is equal to the von Neumann entropy of the
reduced density matrices. This definition is extended to mixed
states through the convex hull construction:
\[
 E(\rho) = {\mathrm{min}} \left\{ \sum_r p_r E(|\psi_r\rangle)
 \; \Big| \; \rho = \sum_r p_r |\psi_r\rangle\langle\psi_r| \right\},
\]
where the minimum is again taken over all possible convex
decompositions of $\rho$.

An analytical lower bound for the entanglement of formation has
been constructed in Ref.~\cite{ALBEVERIO2}, which may be described
as follows. First, for $1 \leq \Lambda \leq N$ one defines the
function
\[
 R(\Lambda) = \min_{\bm\alpha} \left\{ H(\bm{\alpha})
 \; \Big| \; \sum_{ij} \alpha_i \alpha_j = \Lambda \right\}.
\]
The minimum is taken over all Schmidt vectors $\bm{\alpha}$,
i.~e., $R(\Lambda)$ is the minimal value of the entropy
$H(\bm{\alpha})$ under the constraint
$\sum_{ij}\alpha_i\alpha_j=\Lambda$. The solution of this
minimization problem has been derived by Terhal and Vollbrecht
\cite{TERHAL-VOLLBRECHT}:
\[
 R(\Lambda) = H_2(\gamma(\Lambda)) + [1-\gamma(\Lambda)] \log (N-1),
\]
where
\[
 \gamma(\Lambda) = \frac{1}{N^2} \left[\sqrt{\Lambda}
 + \sqrt{(N-1)(N-\Lambda)} \right]^2,
\]
and
\[
 H_2(x) = -x\log x - (1-x) \log (1-x)
\]
is the binary entropy. Second, one introduces the convex hull
${\mathrm{co}}[R(\Lambda)]$ of $R(\Lambda)$. This is the largest
convex function which is bounded from above by $R(\Lambda)$. One
then gets the following lower bound for the entanglement of
formation:
\begin{eqnarray} \label{BOUNDS-EOF-1}
 E(\rho) &\geq& {\mathrm{co}}[R(\Lambda_0)], \nonumber \\
 \Lambda_0 &\equiv& {\mathrm{max}} \left\{
 ||T_2\rho||, ||{\mathcal{R}}\rho|| \right\}.
\end{eqnarray}

The decisive point of the construction given in
Ref.~\cite{ALBEVERIO2} is the fact that [see
Eq.~(\ref{PPT-REALIGN})]
\[
 ||T_2(|\psi\rangle\langle\psi|)||
 = ||{\mathcal{R}}(|\psi\rangle\langle\psi|)||
 = \sum_{ij} \alpha_i \alpha_j = \Lambda.
\]
This means that the function $R(\Lambda)$ yields the minimal
entropy $H(\bm{\alpha})$ under the constraint of a fixed value
$\Lambda$ for the trace norm $||T_2(|\psi\rangle\langle\psi|)||$
or $||{\mathcal{R}}(|\psi\rangle\langle\psi|)||$. But from
Eq.~(\ref{INEQ-FW}) we also have
\[
 1-\langle\psi|W|\psi\rangle \leq \sum_{ij} \alpha_i \alpha_j =
 \Lambda.
\]
By use of this inequality one can immediately repeat the proof of
Ref.~\cite{ALBEVERIO2}, replacing the trace norm $||T_2\rho||$ or
$||{\mathcal{R}}\rho||$ by the quantity $1-{\mathrm{tr}}(W\rho)$.
Hence, we are led to a sharper bound for the entanglement of
formation:
\begin{eqnarray} \label{BOUNDS-EOF-2}
 E(\rho) &\geq& {\mathrm{co}}[R(\Lambda_0)], \nonumber \\
 \Lambda_0 &\equiv& {\mathrm{max}} \left\{
 ||T_2\rho||, ||{\mathcal{R}}\rho||,1-{\mathrm{tr}}(W\rho)
 \right\}.
\end{eqnarray}

\begin{figure}[htb]
\includegraphics[width=\linewidth]{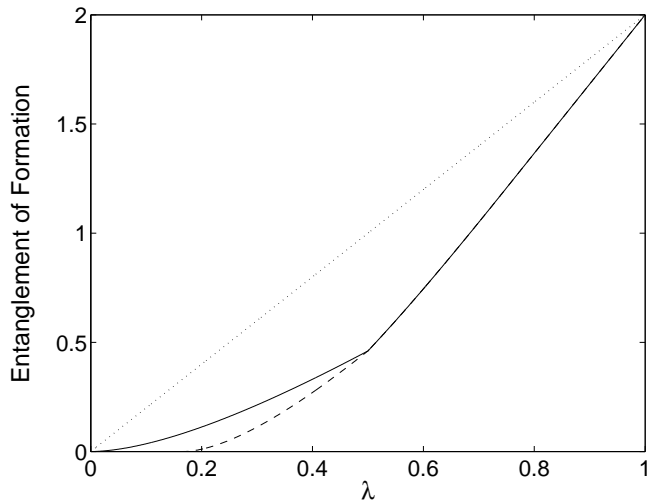}
\caption{Entanglement of formation of the states (\ref{FAMILY})
for $N=4$. Solid line: The new lower bound given by
Eqs.~(\ref{BOUNDS-EOF-2})-(\ref{EOF-BOUND2}). Dashed line: Lower
bound given by Eq.~(\ref{BOUNDS-EOF-1}). Dotted line: Upper bound
given by $E(\rho(\lambda))\leq \lambda \log N$.\label{figure2}}
\end{figure}

Let us apply this result to the family of states given in
Eq.~(\ref{FAMILY}). In this case we have:
\[
 \Lambda_0 = \max \left\{ N\lambda, (N-2)\lambda + 1 \right \}.
\]
By use of the Terhal-Vollbrecht conjecture
\cite{TERHAL-VOLLBRECHT} on the form of the function
${\mathrm{co}}[R(\Lambda)]$ (see also Ref.~\cite{FEI}) we get:
\begin{equation} \label{EOF-BOUND1}
 E(\rho(\lambda)) \geq H_2(\gamma(\Lambda_0))
 + [1-\gamma(\Lambda_0)] \log (N-1)
\end{equation}
for $1\leq\Lambda_0\leq 4(N-1)/N$, and
\begin{equation} \label{EOF-BOUND2}
 E(\rho(\lambda)) \geq
 \frac{\log(N-1)}{N-2}(\Lambda_0-N) + \log N
\end{equation}
for $4(N-1)/N\leq\Lambda_0\leq N$. The general features of this
result are similar to those discussed within the context of the
concurrence. The special case $N=4$ is plotted in
Fig.~\ref{figure2}. We finally note that
\[
 E(\rho(\lambda)) \sim \lambda\log N
\]
represents the asymptotic limit of the entanglement of formation
for large $N$.

\section{Conclusions}\label{CONCLU}

By use of a universal positive map $\Phi$ we have obtained a class
of nondecomposable optimal entanglement witnesses $W$. Employing
these witnesses analytical bounds for the concurrence and for the
entanglement of formation have been developed. Similar bounds can
be derived for other measures, e.~g., for the entanglement measure
which is known as tangle \cite{CAVES}. Due to the fact that $W$ is
a nondecomposable optimal entanglement witness, the bounds
obtained here are particularly good near the boundary which
separates the region of classically correlated states from the
region of entangled states with positive partial transposition.

It should be clear from the general considerations in
Secs.~\ref{SEC-CONCURRENCE} and \ref{SEC-EOF} and from the example
of Sec.~\ref{SEC-EXAMPLE} that the bounds derived here are not
intended to \textit{replace} other known bounds, but rather to
\textit{complement} these. In fact, the bounds based on the
witness $W$ can be weaker than those given by PPT or the
realignment criterion, in particular in those cases in which the
optimal decomposition of $\rho$ consists entirely of maximally
entangled states. To give an example, we consider the family of
states which are invariant under all unitary product
transformations of the form $U\otimes U^*$, were $U^*$ denotes the
complex conjugation of $U$. These states, known as isotropic
states \cite{HORODECKI99}, can be parameterized by a single
parameter, namely by their fidelity $f \in[0,1]$. The isotropic
states for $f\leq 1/N$ are separable and for $f > 1/N$ their
concurrence is given by \cite{CAVES}
\begin{equation} \label{C-F-1}
 C(f) = \sqrt{\frac{2N}{N-1}}\left( f-\frac{1}{N} \right).
\end{equation}
The right-hand side of this equation coincides with the bound
given by the PPT criterion, i.~e., the application of the latter
already yields the exact expression for the concurrence
\cite{ALBEVERIO1}. On the other hand, the bound given by
Eq.~(\ref{NEWBOUND-CONCU}) yields
\begin{equation} \label{C-F-2}
 C(f) \geq \sqrt{\frac{2N}{N-1}} \frac{N-2}{N-1}
 \left( f-\frac{1}{N} \right).
\end{equation}
Since the right-hand side is strictly larger than zero for $f >
1/N$ we can conclude that, like the PPT criterion, also the new
criterion (\ref{TRC}) provides a necessary and sufficient
condition for the separability of the isotropic states. But if we
compare Eqs.~(\ref{C-F-1}) and (\ref{C-F-2}) we see that the bound
obtained by the witness $W$ is always weaker than the bound of the
PPT criterion. Note however that for large $N$ the difference
between these bounds is only of order $1/N$.

We finally indicate some generalizations of the present approach.
An obvious extension concerns the definition of the entanglement
witness $W$ given by Eq.~(\ref{DEF-W}). According to this
definition the witness $W$ depends of course on the chosen basis
and is not invariant under local unitary transformations. However,
for any product transformation $U=U_1 \otimes U_2$ with unitary
operators $U_1$ and $U_2$ the observable $W_U=UWU^{\dagger}$ is
again an entanglement witness. It is clear from the proof given in
Appendix \ref{APP-A} that for arbitrary $U_1$ and $U_2$ the
witness $W_U$ also satisfies the inequality (\ref{INEQ-FW}). It
follows that we may perform the replacement
\[
 {\mathrm{tr}}(W\rho) \longrightarrow
 \min_{U}\{{\mathrm{tr}}(W_U\rho)\}
\]
in the bounds for the concurrence [Eq.~(\ref{NEWBOUND-CONCU})] and
for the entanglement of formation [Eq.~(\ref{BOUNDS-EOF-2})],
where the minimum is taken over all product unitaries $U=U_1
\otimes U_2$. This replacement sharpens the lower bounds and
ensures that they are invariant under local unitary operations.

For simplicity we have restricted ourselves to the case of
bipartite systems with state space ${\mathbb C}^N \otimes {\mathbb
C}^N$. The definition $W=N(I\otimes \Phi)P_0$ for the witness $W$
can be extended to state spaces ${\mathbb C}^M \otimes {\mathbb
C}^N$ with arbitrary $M>N$ in an obvious way, replacing the
singlet state $P_0$ by any maximally entangled pure state in
${\mathbb C}^M \otimes {\mathbb C}^N$. Moreover, an extension of
the present approach to state spaces with odd dimension $N$ seems
to be possible. To this end one has to drop the condition that the
operator $V$ introduced in Sec.~\ref{TIME-REVERSAL} is unitary,
i.~e., one only requires that $V$ is skew-symmetric and that
$V^{\dagger}V \leq I$. It is worth investigating applications of
these constructions to bipartite and multipartite quantum systems.

\appendix

\section{Proof of inequality (\ref{INEQ-AIJ})}\label{APP-A}

To prove inequality (\ref{INEQ-AIJ}) we consider a fixed pair
$(i,j)$ of indices and decompose $|\theta\chi_i\rangle$ into a
component which is parallel to $|\chi_j\rangle$ and a component
which is perpendicular to $|\chi_j\rangle$:
\begin{equation} \label{APP-A-1}
 |\theta\chi_i\rangle = \lambda |\chi_j\rangle +
 \mu |\chi_j^{\perp}\rangle,
\end{equation}
where $\langle \chi_j | \chi_j^{\perp}\rangle = 0$ and
$|\lambda|^2+|\mu|^2=1$. Applying the time reversal transformation
to this equation we get:
\begin{equation} \label{APP-A-2}
 |\chi_i\rangle = -\lambda^* |\theta\chi_j\rangle
 -\mu^* |\theta\chi_j^{\perp}\rangle.
\end{equation}
Inserting (\ref{APP-A-1}) and (\ref{APP-A-2}) into the expression
(\ref{DEF-AIJ}) one obtains:
\[
 A_{ij} = \mu \left[
 \langle\varphi_i|\chi_j\rangle \langle\theta\varphi_j|\chi_j^{\perp}\rangle -
 \langle\varphi_i|\chi_j^{\perp}\rangle \langle\theta\varphi_j|\chi_j\rangle
 \right].
\]
Since $|\mu|\leq 1$ this leads to
\[
 |A_{ij}| \leq ab + cd,
\]
where we have introduced the quantities:
\begin{eqnarray*}
 a = |\langle\varphi_i|\chi_j\rangle|, && \qquad
 b = |\langle\theta\varphi_j|\chi_j^{\perp}\rangle|, \\
 c = |\langle\varphi_i|\chi_j^{\perp}\rangle|, && \qquad
 d = |\langle\theta\varphi_j|\chi_j\rangle|.
\end{eqnarray*}
Since $|\chi_j\rangle$ is perpendicular to
$|\chi_j^{\perp}\rangle$ by construction, we have
\[
 a^2 + c^2 = |\langle\varphi_i|\chi_j\rangle|^2 +
 |\langle\varphi_i|\chi_j^{\perp}\rangle|^2 \leq 1,
\]
and, therefore, $c \leq \sqrt{1-a^2}$. In a similar manner we get
$d \leq \sqrt{1-b^2}$. Hence, we obtain:
\[
 |A_{ij}| \leq ab + \sqrt{1-a^2}\sqrt{1-b^2}.
\]
It is easy to see that the right-hand side of this inequality is
smaller than or equal to 1 for all $a,b \in [0,1]$, which yields
the desired inequality (\ref{INEQ-AIJ}).

\section{Determination of the trace norms $||T_2\rho(\lambda)||$
         and $||{\mathcal{R}}\rho(\lambda)||$}\label{APP-B}

Since $T_2\rho$ and $\vartheta_2\rho$ are unitarily equivalent we
have $||T_2\rho(\lambda)||=||\vartheta_2\rho(\lambda)||$. Using
Eqs.~(\ref{FAMILY}) and (\ref{THETA-P0}), and the representation
$P_S=(I+F)/2$ we get
\[
 \vartheta_2 \rho(\lambda) = \frac{1-2\lambda}{N}P_0
 + \frac{1}{N} \sum_{J=1}^{2j}
 \left[ (-1)^{J+1} \lambda + \frac{1-\lambda}{N+1} \right] P_J.
\]
Hence, the trace norm of $T_2\rho(\lambda)$ is found to be:
\[
 ||T_2 \rho(\lambda)|| = \frac{|1-2\lambda|}{N}
 + \sum_{J=1}^{2j} \frac{2J+1}{N}
 \left| (-1)^{J+1} \lambda + \frac{1-\lambda}{N+1} \right|.
\]
Carrying out this sum one gets:
\[
 ||T_2 \rho(\lambda)|| - 1 =
 \left\{
 \begin{array}{ll}
 0, & \lambda \leq \frac{1}{N+2} \\
 \frac{N-2}{N} \left[ (N+2)\lambda - 1 \right],
 & \frac{1}{N+2} \leq \lambda \leq \frac{1}{2} \\
 N\lambda-1, & \frac{1}{2} \leq \lambda
 \end{array} \right.
\]
which yields the lower bounds of Eq.~(\ref{PPT-BOUND}).

To determine $||{\mathcal{R}}\rho(\lambda)||$ we note that the
realignment transformation may be written as ${\mathcal{R}}\rho =
\vartheta_2(F\rho)$. Using $FP_0=-P_0$ and $FP_S=P_S$ one easily
deduces that
\[
 {\mathcal{R}}\rho(\lambda) = \frac{1}{N}P_0
 + \frac{1}{N} \sum_{J=1}^{2j}
 \left[ (-1)^J \lambda + \frac{1-\lambda}{N+1} \right] P_J,
\]
which yields:
\[
 ||{\mathcal{R}}\rho(\lambda)|| = \frac{1}{N}
 + \sum_{J=1}^{2j} \frac{2J+1}{N}
 \left| (-1)^J \lambda + \frac{1-\lambda}{N+1} \right|.
\]
The evaluation of this sum leads to:
\[
 ||{\mathcal{R}}\rho(\lambda)|| - 1 = \left\{
 \begin{array}{ll}
 -2\lambda, & \lambda \leq \frac{1}{N+2} \\
 N\lambda-1, & \frac{1}{N+2} \leq \lambda
 \end{array} \right.
\]
which gives the lower bounds of Eq.~(\ref{REALIGN-BOUND}).

\end{document}